\documentclass[aps,prb,twocolumn,showpacs,superscriptaddress]{revtex4}

\usepackage{graphicx} 

\begin{document}

\title
{Hydrogen Absorption Properties of Metal-Ethylene Complexes}

\author{W. Zhou}
\affiliation{NIST Center for Neutron Research, National Institute
of Standards and Technology, Gaithersburg MD 20899}
\affiliation{Department of Materials Science and
Engineering, Univ. of Pennsylvania, Philadelphia, PA 19104}
\author{T. Yildirim}
\email{taner@nist.gov} 
\affiliation{NIST Center for Neutron Research, National Institute
of Standards and Technology, Gaithersburg MD 20899}
\affiliation{Department of Materials Science and
Engineering, Univ. of Pennsylvania, Philadelphia, PA 19104}
\author{E. Durgun}
\affiliation{Department of Physics, Bilkent University, Ankara
06800, Turkey} \affiliation{UNAM - National Nanotechnology
Research Center, Bilkent University, Ankara 06800, Turkey}
\author{S. Ciraci}
\affiliation{Department of
Physics, Bilkent University, Ankara 06800, Turkey}
\affiliation{UNAM - National Nanotechnology Research Center,
Bilkent University, Ankara 06800, Turkey}

\date{\today}

\begin{abstract}

Recently, we have predicted [Phys. Rev. Lett. {\bf 97}, 226102 (2006)] 
that a single ethylene molecule can form stable complexes with 
light transition metals (TM) such as Ti and the resulting 
TM$_n$-ethylene complex can absorb up to  $\sim$12 and 14 wt \% hydrogen 
for n=1 and 2, respectively. Here we extend this study to 
include a large number of other metals and different isomeric structures. We obtained interesting 
results for light metals such as Li. The ethylene molecule is 
able to complex with two Li atoms with a binding energy of 0.7 eV/Li
which then binds up to two H$_2$ molecules per Li with a binding energy of 0.24 eV/H$_2$
and absorption capacity of 16 wt \%, a record high value reported so far.
The stability of the proposed metal-ethylene complexes was tested by extensive calculations  
such as normal-mode analysis,  finite temperature first-principles 
molecular dynamics (MD) simulations, and  reaction path 
calculations. The phonon and MD simulations indicate
that the proposed structures are stable up to 500 K. The
reaction path calculations indicate about 1 eV activation
barrier for the TM$_2$-ethylene complex to transform into a possible
lower energy configuration where the ethylene molecule is 
dissociated. Importantly, no matter which isometric configuration the
TM$_2$-ethylene complex possesses, the TM atoms are able to bind multiple
hydrogen molecules with suitable binding energy for room temperature storage.
These results suggest that co-deposition of ethylene with
a suitable precursor of TM or Li into nanopores of 
light-weight host materials may be a very promising route to discovering 
new materials with high-capacity hydrogen absorption properties. 
\end{abstract}

\pacs{68.43.Bc, 81.07.-b,84.60.Ve}

\maketitle

\section{Introduction}

The success of future hydrogen and fuel-cell technologies is critically 
dependent upon the discovery of new materials that can store a large amount 
of hydrogen at ambient 
conditions.\cite{See:2004}$^{,}$\cite{Crabtree:2004}$^{,}$\cite{Zuttel:2003} 
Recently, from quantum mechanical calculations we found that the C=C bond in 
a single ethylene molecule, similar to C$_{60}$ and carbon 
nanotubes\cite{Yildirim:2005}$^{,}$\cite{Yildirim:2006}$^{,}$\cite{Zhao:2005}$^{,}$\cite{Dag:2005}$^{,}$\cite{Kiran:2006}, 
can form a stable complex with transition metals (TM) such as 
Ti.\cite{Durgun:2006} The resulting TM$_{2}$-ethylene complex attracts up 
to ten hydrogen molecules via the Dewar-Kubas interaction\cite{Metal:2001}, 
reaching a gravimetric storage capacity of ~$\sim $14 weight-percent (wt 
{\%}).\cite{Durgun:2006} The interaction between hydrogen molecules and 
transition metals lies between chemi- and physi- sorption, with a binding 
energy of $\sim $0.4 eV/H$_{2}$ compatible with room temperature 
desorption/absorption. 

Different from metal decorated C$_{60}$/nanotubes, metal-C$_{2}$H$_{4}$ 
complexes are actually existing structures and have been actively studied in 
the past several decades, with the major goal being to understand the 
catalytic mechanisms and processes of metals. Experimental spectroscopic 
data on various complexes, such as Li, Mg, Al and TMs complexed with 
C$_{2}$H$_{4}$, widely exist in the 
literature\cite{Manceron:1986}$^{,}$\cite{Chen:1998}$^{,}$\cite{Manceron:1989}$^{,}$\cite{Ozin:1978}. 
These complexes were typically synthesized by direct reaction of metal atoms 
with C$_{2}$H$_{4}$/Ar in the gas phase. Early theoretical 
studies\cite{Ozin:1978}$^{,}$\cite{Alikhani:1996}$^{,}$\cite{Sodupe:1992}$^{,}$\cite{Blomberg:1992}$^{,}$\cite{Papai:1993} 
showed that the metal-C$_{2}$H$_{4}$ binding mechanisms could be either 
electrostatic (e.g., C$_{2}$H$_{4}$-Al), or Dewar-Chatt-Duncanson bonding 
(e.g., most C$_{2}$H$_{4}$-TMs). Despite the extensive studies on the 
metal-C$_{2}$H$_{4}$ complexes, their ability to absorb H$_{2}$ was not 
realized and investigated until our recent work\cite{Durgun:2006}. 

Here we extend our earlier work\cite{Durgun:2006} and present a detailed 
theoretical study of the hydrogen absorption on a large number of 
metal-C$_{2}$H$_{4}$ complexes, including TMs and the alkali metal Li. We 
organize the paper as follows. In the next section, we describe the 
computational methodology. In Sec.III, we discuss C$_{2}$H$_{4}$M complexes, 
various isomers of C$_{2}$H$_{4}$M$_{2}$ complexes and present the metal 
binding energies, zero-temperature lattice dynamics of these complexes and 
their hydrogen absorption properties (including the H$_{2}$ binding energies 
and maximum number of H$_{2}$ that the complex can absorb). In Sec. IV, we 
discuss the possible reaction paths (i.e., minimum energy paths) and the 
activation energies (i.e. barriers) between various isomers of 
C$_{2}$H$_{4}$Ti$_{2}$ complexes. We also discuss an interesting catalytic 
effect of Ti, similar to the ``spillover effect'', where a molecularly bound 
H$_{2}$ molecule is first dissociated over Ti and then one of the H atoms is 
bonded to carbon, forming a CH$_{3}$ group. The resulting molecule is 
isostructural to an ``ethanol'' molecule and thus called ``titanol''. The 
titanol molecule is also able to absorb up to five H$_{2}$ as molecules with 
a binding energy of $\sim $0.4 eV/H$_{2}$ and provide another interesting 
possibility for high-capacity hydrogen storage materials. In Sec.V, we 
present high-temperature first-principles molecular dynamics (MD) studies on 
selected structures. Due to the small system size, we are able to carry out 
MD simulations up to 10 ps. We show that the proposed complex structures are 
quite stable and exhibit constructive desorption upon heating without 
destroying the underlying complex. Our concluding remarks are presented in 
Sec. VI.

\section{Details of calculations}

Our first-principles energy calculations were done within density functional 
theory using Vanderbilt-type ultra-soft pseudopotentials with Perdew-Zunger 
exchange correlation, as implemented in the PWscf package.\cite{Baroni:1} 
Single molecular complexes have been treated in a supercell of 20$\times 
$20$\times $20 {\AA}~with $\Gamma$ k-point and a cutoff energy of 408 eV. The 
structures are optimized until the maximum force allowed on each atom is 
less than 0.01 eV/ {\AA} for both spin-paired and spin-relaxed cases. The 
reaction path calculations were carried out using the Nudged Elastic Band 
(NEB) method\cite{Mills:1994}$^{,}$\cite{Henkelman:2000}. We used a total 
of 21 images between the reactant and the product, which were fully 
optimized during the NEB calculations. The MD simulations were carried out 
within the microcanonical ensemble (NVE) starting with the optimized 
structure and random initial atom 
velocities.\cite{Marx:2000}$^{,}$\cite{Frenkel:1996} More details of the 
MD calculations are given in Sec.V.

\section{Structural, Electronic, and Dynamical Properties of C$_{2}$H$_{4}$M$_{n}$ 
and C$_{2}$H$_{4}$M$_{n}$-H$_{x}$ complexes}

\begin{table*}[htbp]
\caption{The metal-C$_{2}$H$_{4}$ binding energies (in eV/M-atom) with respect to atomic and bulk 
energies of various metals, and the average H$_{2}$ binding energies (in 
eV/H$_{2}$) on C$_{2}$H$_{4}$M for various absorption configurations (see Fig. 3). The maximum number of H$_{2}$ molecules bonded to 
each metal is also shown.}
\begin{center}
\begin{tabular}{c|cccccccccccccccc}
\hline
Property/M & 
Li& 
Sc & 
Ti & 
V & 
Cr & 
Mn & 
Fe & 
Co & 
Ni & 
Cu & 
Zn & 
Zr & 
Mo & 
W & 
Pd & 
Pt  \\
\hline
$E_{B}$ (M-atomic)& 
0.32& 
1.39& 
1.45& 
0.94& 
0.18& 
0.51& 
0.92& 
1.39& 
0.91& 
0.80& 
none & 
1.91& 
1.02& 
1.71& 
1.95& 
2.52 \\
$E_{B}$ (M-bulk)& 
-1.41& 
-2.72& 
-3.68& 
-4.30& 
-3.44& 
-3.06& 
-1.64& 
-2.43& 
-1.99& 
-2.86& 
--& 
-4.23& 
-5.19& 
-6.65& 
-1.86& 
-2.82 \\
$E_{B}$ (per H$_{2})$, MH$_{2}$& 
--& 
0.96& 
1.16& 
1.00& 
0.01& 
0.59& 
1.01& 
0.94& 
1.13& 
0.19& 
--& 
1.90& 
0.85& 
1.80& 
0.83& 
1.33 \\
$E_{B}$ (per H$_{2})$, M+H$_{2}$& 
0.29& 
0.02& 
0.31& 
0.46& 
0.45& 
--& 
--& 
--& 
--& 
--& 
--& 
0.35& 
0.49& 
0.59& 
0.64& 
-- \\
max H$_{2}$/M& 
2& 
5 & 
5 & 
5 & 
5 & 
5 & 
5 & 
3 & 
2 & 
2 & 
-- & 
5 & 
5 & 
5 & 
2 & 
2  \\
$E_{B}$ (per H$_{2})$, M+2H$_{2}$& 
0.28& 
--& 
--& 
--& 
--& 
--& 
--& 
--& 
0.42& 
0.33& 
--& 
--& 
--& 
--& 
0.27& 
0.25 \\
$E_{B}$ (per H$_{2})$, MH$_{2}$+3H$_{2}$& 
--& 
0.40& 
0.54& 
0.66& 
0.34& 
0.24& 
0.31& 
--& 
--& 
--& 
--& 
0.78& 
0.61& 
0.85& 
--& 
-- \\
$E_{B}$ (per H$_{2})$, M+5H$_{2}$& 
--& 
0.28& 
0.46& 
0.53& 
0.21& 
0.18& 
0.34& 
--& 
--& 
--& 
--& 
0.54& 
0.64& 
0.79& 
--& 
-- \\
\hline
\end{tabular}
\label{tab1}
\end{center}
\end{table*}

\begin{table*}[htbp]
\caption{The metal-C$_{2}$H$_{4}$ binding energies (in eV/M-atom) of three isomeric 
C$_{2}$H$_{4}$M$_{2}$ configurations (see Fig. 1), with respect to atomic and dimer 
energies of various metals. }
\begin{center}
\begin{tabular}{c|cccccccccccccccc}
\hline
Property/M & 
Li& 
Sc & 
Ti & 
V & 
Cr & 
Mn & 
Fe & 
Co & 
Ni & 
Cu & 
Zn & 
Zr & 
Mo & 
W & 
Pd & 
Pt  \\
\hline
$E_{B}$ (M-atomic), Sandwich& 
0.69& 
1.39 & 
1.47 & 
1.21 & 
0.05 & 
0.37 & 
0.83 & 
1.30 & 
0.70 & 
1.41 & 
none & 
1.69 & 
0.37 & 
1.18 & 
1.56 & 
1.78  \\
$E_{B}$ (M-atomic), Dimer-par& 
0.54& 
1.77& 
2.02& 
1.62& 
0.10& 
0.64& 
1.65& 
1.63& 
1.09& 
1.34& 
none& 
2.66& 
2.20& 
3.26& 
1.88& 
2.61 \\
$E_{B}$ (M-atomic), Dimer-perp& 
0.61& 
1.72& 
2.12& 
1.97& 
0.21& 
0.51& 
1.22& 
1.50& 
1.00& 
1.24& 
none& 
2.70& 
2.10& 
2.41& 
1.39& 
1.53 \\
$E_{B}$ (M-dimer), Sandwich& 
0.20& 
0.58& 
0.17& 
-0.21& 
0.79& 
0.33& 
-0.36& 
-3.45& 
-0.01& 
0.17& 
--& 
-0.10& 
-1.71& 
-1.30& 
0.75& 
0.20 \\
$E_{B}$ (M-dimer), Dimer-par& 
0.05& 
0.96& 
0.72& 
0.20& 
0.84& 
0.60& 
0.46& 
-3.12& 
0.38& 
0.10& 
--& 
0.87& 
0.12& 
0.78& 
1.08& 
1.03 \\
$E_{B}$ (M-dimer), Dimer-perp& 
0.12& 
0.91& 
0.82& 
0.54& 
0.95& 
0.47& 
0.03& 
-3.25& 
0.29& 
0.00& 
--& 
0.91& 
0.02& 
-0.07& 
0.58& 
-0.05 \\
\hline
\end{tabular}
\label{tab2}
\end{center}
\end{table*}

We start by examining various possible configurations of 
C$_{2}$H$_{4}$M$_{n}$ complexes and their corresponding H$_{2}$ absorption 
properties. We consider both transition metals and light metal Li, and focus 
on $n$=1 and $n$=2 cases. Complexes with $n>$2 are less attractive for hydrogen 
storage due to potentially lower capacities and thus are not discussed here 
and should be avoided in the syntheses.

\begin{figure}
\includegraphics[width=7cm]{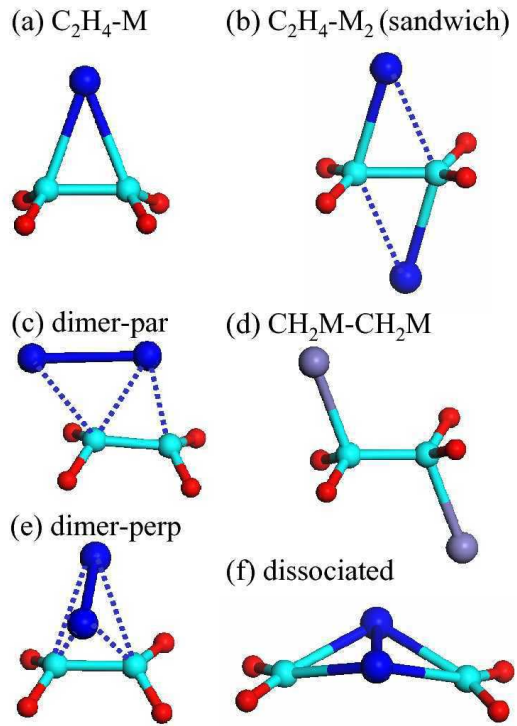}
\caption{
(Color online) Various configurations of C$_{2}$H$_{4}$M$_{n}$ ($n$=1 and 2) 
complexes considered in this study. (a) C$_{2}$H$_{4}$ complexed 
with one metal atom. (b)-(e) C$_{2}$H$_{4}$ complexed with two metal atoms 
with different metal binding sites. Note that the bond-stick model is only used for 
clarity and should not be considered as an implication of the chemical 
covalent bonding between those atoms. For most metals, there is no classical 
chemical covalent bonding between the metal and carbon atoms. For a few 
metals (e.g., Fe), the complexes possess a structure, where 
M and C are bonded more traditionally by covalent bonding, as shown in (d). (f) 
C$_{2}$H$_{4}$M$_{2}$ complex with dissociated C=C bond. Large, medium and small
balls represent M, C and H atoms, respectively.
}
\label{c2h4mx-config}
\end{figure}

When one metal atom binds to the ethylene molecule, the configuration shown 
in Fig. 1(a) is the most energetically favorable one, where the metal atom 
forms a symmetric bridge ``bond'' with the C=C bond of ethylene. When two 
metal atoms bind to C$_{2}$H$_{4}$, the complex may adopt several possible 
configurations. In our initial study\cite{Durgun:2006}, we focused on 
the sandwich structure (Fig. 1(b)). Here we consider two additional isomeric 
structures: dimer-par (Fig. 1(c)) and dimer-perp (Fig. 1(e)). In the 
sandwich configuration, each M atom is closer to one of the carbon atoms, 
leading to two different M-C ``bonds''. Note that for most transition metals 
(e.g., Ti), there is no classical chemical covalent bonding between the 
metal atom and carbon atom. The calculated bond-population is found to be 
nearly zero for these metals. The slight shift of the metal atoms towards 
different C atoms only results in a minute contribution of the M-C 
covalent-like bond to the overall binding. In just a few cases (e.g., Fe), 
the metal and carbon atom are bonded more traditionally by a covalent bond, 
as shown in Fig. 1(d). For this reason, we generally specify these 
C$_{2}$H$_{4}$M$_{n}$ structures as "complexes" instead of "molecules".

The binding mechanism of the C$_{2}$H$_{4}$TM complex has been discussed in 
detail in our previous work\cite{Durgun:2006}. Essentially, the bonding 
orbital for the TM-atoms and C$_{2}$H$_{4}$ results from the hybridization 
of the lowest-unoccupied molecular orbital (LUMO) of the ethylene molecule 
and the TM-$d$ orbitals, in accord with Dewar coordination. For Li, the binding 
mechanism is, however, different. In Fig. 2, we show the electronic density 
of states of C$_{2}$H$_{4}$ molecule, Li atom and C$_{2}$H$_{4}$Li complex. 
Projection analysis of the states indicates that the electron in the 2s 
state of Li is divided into two halves that are transferred to the LUMO of 
C$_{2}$H$_{4}$ and the 2p of the Li atom, respectively. Then the 2p orbital 
of Li and the LUMO of C$_{2}$H$_{4}$ are hybridized for the binding of Li on 
the C$_{2}$H$_{4}$. From the isosurfaces of the molecular orbitals (also 
shown in Fig. 2), it is clear that the molecular orbital of C$_{2}$H$_{4}$Li 
near the zero-energy (i.e., the Fermi Energy), is a superposition of the 
LUMO of C$_{2}$H$_{4}$ and the p-orbital of the Li atom. Also note that the 
occupied orbital of the C$_{2}$H$_{4}$Li complex at around -4eV is about 
the same as that the highest occupied molecular orbital (HOMO) of bare 
C$_{2}$H$_{4}$, except that there is a hole in the upper portion of the 
orbital due to the Li-ion. The bond analysis does not show any covalent 
bonding between C and Li atoms. For C$_{2}$H$_{4}$Li$_{2}$, we observed also 
a binding mechanism similar to that of C$_{2}$H$_{4}$Li.

The metal binding energies on ethylene are summarized in Table I and Table 
II for one metal and two metal complexes, respectively. They are calculated 
by subtracting the equilibrium total energy $E_{T}$ of the 
C$_{2}$H$_{4}$M$_{n}$ complex from the sum of the total energies of free 
molecular ethylene and of the M atom: $E_{B}$(M) = [$E_{T}$(C$_{2}$H$_{4})$ + 
\textit{nE}$_{T}$(M) - $E_{T}$(C$_{2}$H$_{4}$M$_{n})$]/$n$. According to the 
$E_{B}$(M-atomic) values shown in the table, most TMs that we studied are 
able to bind relatively strongly to a C$_{2}$H$_{4}$ molecule, except Cr and 
Zn. In table I, the variation of the TM binding energy with the number of 
the TM-3$d$ electrons displays a behavior similar to what observed previously 
for the chemisorption of TMs on the surface of a single-walled carbon 
nanotube\cite{Durgun:2003}$^{,}$\cite{Durgun:2004}. Namely, there exist 
two energy maxima between a minimum that occurs for the element with five 
$d$-electrons. Table~I also gives the binding energies with respect to bulk 
metal energies ($E_{B}$(M-bulk)) while Table II also gives $E_{B}$ with 
respect to metal dimer energies ($E_{B}$(M-dimer)). All $E_{B}$(M-bulk) values 
are negative, indicating endothermic reactions. Apparently, metal atoms in 
vapor or some metal-precursors, instead of bulk metals, are preferred when 
synthesizing these complex structures.

\begin{figure}
\includegraphics[width=7cm]{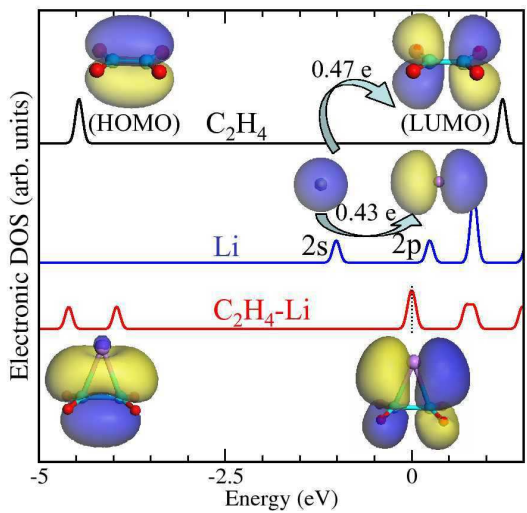}
\caption{
(Color online) Electronic density of states of  C$_{2}$H$_{4}$, Li 
atom, and C$_{2}$H$_{4}$+Li complex. The isosurfaces of 
the molecular orbitals are also shown. 
The hybridization of the Li-2p state and the LUMO of C$_{2}$H$_{4}$
is apparent. See text for further explanation.
}
\label{c2h4lix}
\end{figure}

\begin{figure}
\includegraphics[width=6cm]{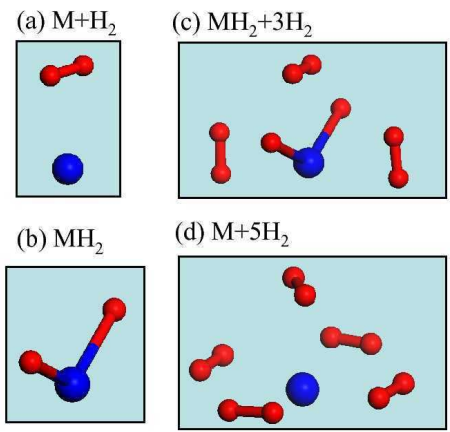} 
\caption{
(Color online) Various configurations that we considered in this study, 
for the hydrogen absorption on a metal center of a C$_{2}$H$_{4}$M$_{n}$ complex: (a) one H$_{2}$ absorbed 
molecularly; (b) H$_{2}$ dissociating with two M-H bond formed; (c) two atomic H 
and three H$_{2}$ molecules; (d) five H$_{2}$ absorbed as molecules. Large and small
balls represent M and H atoms, respectively.
}
\label{mhx-config}
\end{figure}

\begin{figure}
\includegraphics[width=7cm]{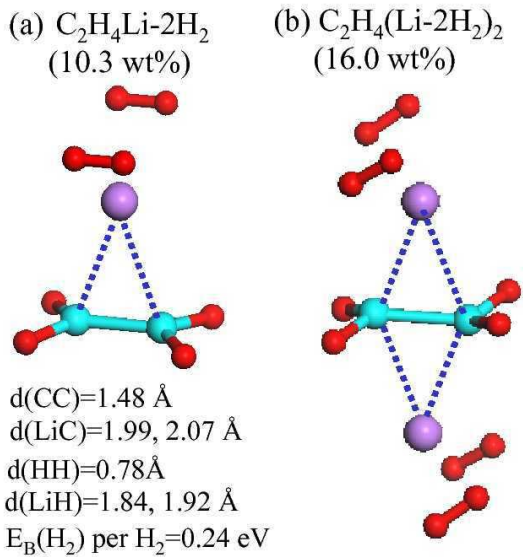}
\caption{
(Color online) Hydrogen absorption configurations on (a) C$_{2}$H$_{4}$Li 
and (b) C$_{2}$H$_{4}$Li$_{2}$ complexes. Note that in both cases, 
each Li can bind two H$_{2}$, resulting in high absorption capacities.
Large, medium and small balls represent Li, C and H atoms, respectively.
}
\label{c2h4-lix-yh2}
\end{figure}

We next studied the H$_{2}$ storage capacity of the metal-ethylene complex, 
by calculating the interaction between C$_{2}$H$_{4}$M$_{n}$ and different 
number of H$_{2}$ molecules. We considered various configurations for the 
hydrogen absorption on a metal center, as shown in Fig. 3. The first H$_{2}$ 
molecule absorbed may either be in molecular form (Fig. 3(a)) or in 
dissociated form (Fig.~3(b)). For most transition metals, it is possible to 
absorb more, up to five H$_{2}$ per M atom. Two of the many possible 
multiple H$_{2}$ absorption configurations are shown in Fig. 3 (c) and (d). 
For Li, in both C$_{2}$H$_{4}$Li and C$_{2}$H$_{4}$Li$_{2}$ complexes, each 
Li can bind to two H$_{2}$, resulting in absorption capacity of 10.3 wt {\%} 
and 16.0 wt {\%} respectively. The optimized configurations and structural 
parameters are shown in Fig. 4.

The nature of the metal-H$_{2}$ interaction is easy to understand. For TMs, 
since the bonding orbitals are mainly between metal $d$- and hydrogen $\sigma 
^{\ast }$-antibonding orbitals, the mechanism of this interesting 
interaction can be explained by the Kubas interaction\cite{Metal:2001}. 
For Li, the metal-H$_{2}$ binding is mainly electrostatic. We summarize the 
average H$_{2}$ binding energy for C$_{2}$H$_{4}$M in Table~I. Note that the 
H$_{2}$ binding energies differ slightly from those given in our earlier 
work for the C$_{2}$H$_{4}$M$_{2}$ complexes with the sandwich structure. 
Nevertheless, in most cases, the H$_{2}$ binding energies have the right 
order of magnitude for room temperature storage. Since the hydrogens are 
mainly absorbed molecularly, we also expect fast absorption/desorption 
kinetics. 

In order to test their stability, we further studied the dynamic of the 
C$_{2}$H$_{4}$M$_{n}$ complexes by normal mode analysis. We found no soft 
(i.e negative) mode, indicating that the complex structures are stable. 
Characteristic phonon modes are summarized in Table III, using Li and Ti as 
examples. Our calculated mode frequencies for the C$_{2}$H$_{4}$ molecule 
agree very well with the experimental values\cite{Georges:1999}. Metal 
binding to C$_{2}$H$_{4}$ elongates and thus softens the C=C bond, resulting 
in lower stretching mode frequencies. Also the softening of the 
CH$_{2}$-torsion and CH$_{2}$-bending modes is obvious. There are three main 
M-related vibrational modes. In two of these modes, M atoms vibrate parallel 
and perpendicular to the C=C bond. In the third mode, metal atoms vibrate 
perpendicular to the C$_{2}$H$_{4}$ plane. These three modes are unique for 
the C$_{2}$H$_{4}$M$_{n}$ complex and therefore should be present in any 
Raman/IR spectra of a successfully synthesized material.

\begin{figure}
\includegraphics[width=7cm]{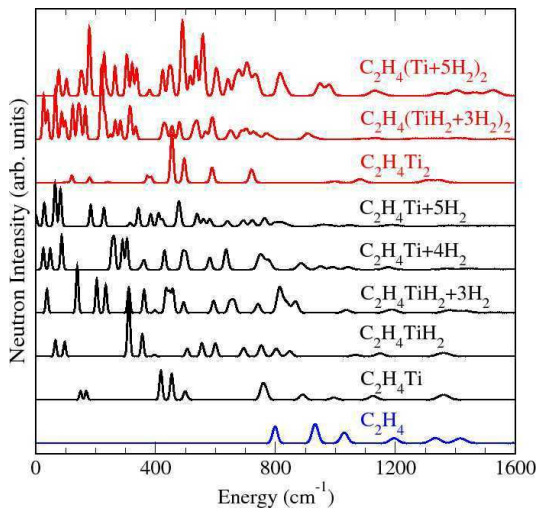} 
\caption{
(Color online) Simulated neutron inelastic spectrum for various 
C$_{2}$H$_{4}$Ti$_{n}$-H$_{x}$ configurations. Note that the M-H dynamics
are unique and can be used as a probe to identify these 
structures. Thus these plots can provide a useful comparison to
experiments when trying to synthesize these materials. 
}
\label{ins}
\end{figure}

\begin{table*}[htbp]
\caption{Characteristic mode frequencies (meV) for C$_{2}$H$_{4}$, 
C$_{2}$H$_{4}$Li$_{n}$ and C$_{2}$H$_{4}$Ti$_{n}$ complexes. Experimental 
values for C$_{2}$H$_{4}$ (from ref. 24) are also shown. Note that the 
metal-C$_{2}$H$_{4}$ binding significantly softens the C=C stretching, 
CH$_{2}$-torsion and CH$_{2}$-bending modes. The three main M-modes give 
unique signatures for metal-C$_{2}$H$_{4}$ complexes.}
\begin{center}
\begin{tabular}{c|cccccc}
\hline
Mode/Complex& 
C$_{2}$H$_{4}$& 
C$_{2}$H$_{4}$, exp.& 
C$_{2}$H$_{4}$Li& 
C$_{2}$H$_{4}$Li$_{2}$& 
C$_{2}$H$_{4}$Ti& 
C$_{2}$H$_{4}$Ti$_{2}$ \\
\hline
C=C stretching& 
202 & 
201& 
184& 
172& 
170& 
167 \\
CH$_{2}$-torsion& 
128 & 
127& 
99& 
49& 
52& 
56 \\
CH$_{2}$-bending& 
115-165 & 
117-166& 
84-145& 
54-141& 
94-140& 
73-134 \\
M-vib, // C=C bond& 
--& 
--& 
38& 
40 (in-phase), \par 66 (out-of-phase)& 
62& 
15 (in-phase), \par 57 (out-of-phase) \\
M-vib, $\bot $ C=C bond& 
--& 
--& 
37& 
22 (in-phase), \par 65 (out-of-phase)& 
56& 
22 (in-phase), \par 62 (out-of-phase) \\
M-vib, $\bot $ C$_{2}$H$_{4}$ plane& 
--& 
--& 
40& 
38 (in-phase), \par 74 (out-of-phase)& 
63& 
29 (in-phase), \par 48 (out-of-phase) \\
\hline
\end{tabular}
\label{tab3}
\end{center}
\end{table*}

We also calculated the normal modes of C$_{2}$H$_{4}$M$_{n}$ complexes 
absorbed with H$_{2}$ and did not find any soft modes either, indicating 
that the configurations that we considered indeed correspond to local energy 
minima. Among many vibrational modes, we note that the H$_{2}$ stretching 
mode is around 330-420 meV for the absorbed H$_{2}$ molecules, significantly 
lower than $\sim $540 meV for the free H$_{2}$ molecule. Such a shift in the 
mode frequency would be the key feature that can be probed by Raman/IR 
measurement to confirm a successful synthesis of the structures predicted 
here. In the lower energy range, there are many M-H modes that are unique to 
the complexes. To manifest the M-H dynamics, we show in Fig. 5 the phonon 
density of states of C$_{2}$H$_{4}$Ti$_{n}$--H$_{x}$ complexes weighted by 
neutron cross-sections (note that H has much larger neutron scattering cross 
section than C and most metals). These plots can provide a useful comparison 
to experiments when trying to synthesize these materials.

\section{Activation Energies and Reaction Paths between Different Isomers}

\begin{figure}
\includegraphics[width=7cm]{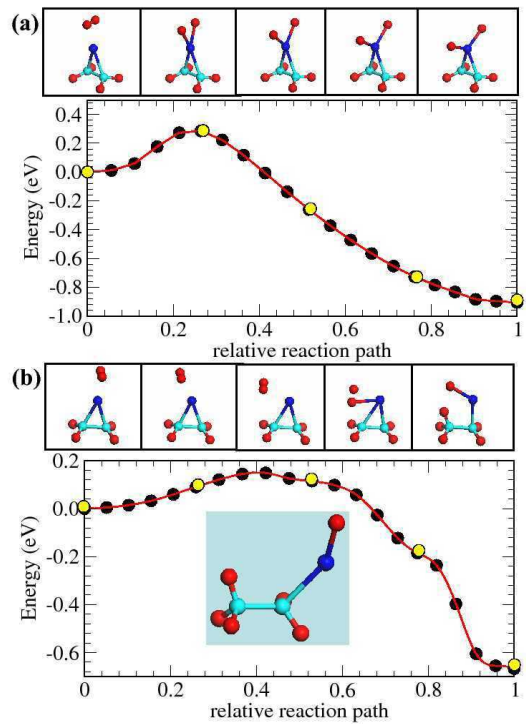}
\caption{
(Color online) (a) The minimum energy path for the dissociation of the 
H$_{2}$ molecule over the Ti atom complexed with C$_{2}$H$_{4}$. An energy 
barrier of $\approx $ 0.25 eV is found for the dissociation. A total of 21 
images were used in the NEB calculations, five of which are shown on the 
top. Marked circles in the potential plot are the points corresponding to 
these five images. (b) The activation energy plot for the formation of 
Titanol-molecule from C$_{2}$H$_{4}$Ti+H$_{2}$ complex, indicating a very low 
barrier of $\approx $0.15 eV. Once the final product forms, the CCTi-bond angle is 
very soft, resulting in the zero-temperature structure shown in the inset.
}
\label{rp_h2_disso}
\end{figure}

\begin{figure}
\includegraphics[width=6cm]{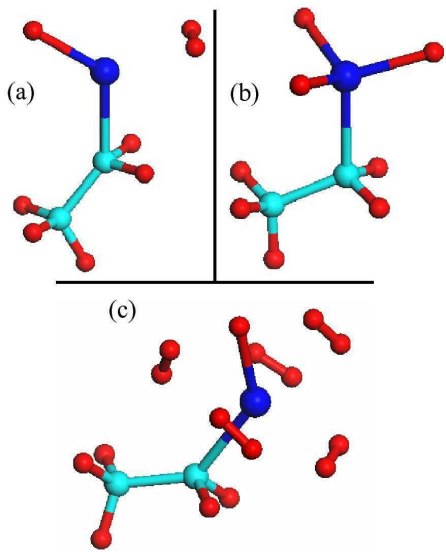}
\caption{
(Color online) Hydrogen absorption on a titanol molecule. (a) One H$_{2}$ 
binds molecularly to Ti, with a binding energy of 0.3 eV. (b) One H$_{2}$ is 
dissociated, yielding TiH$_{3}$ structure. The corresponding binding energy is 
about 1.0 eV/H$_{2}$. (c) Five H$_{2}$ bind as molecules to the 
titanol molecule with an average binding energy of 0.4 eV/H$_{2}$.
Large, medium and small balls represent Ti, C and H atoms, respectively.
}
\label{rp_titanol}
\end{figure}

\begin{figure}
\includegraphics[width=7cm]{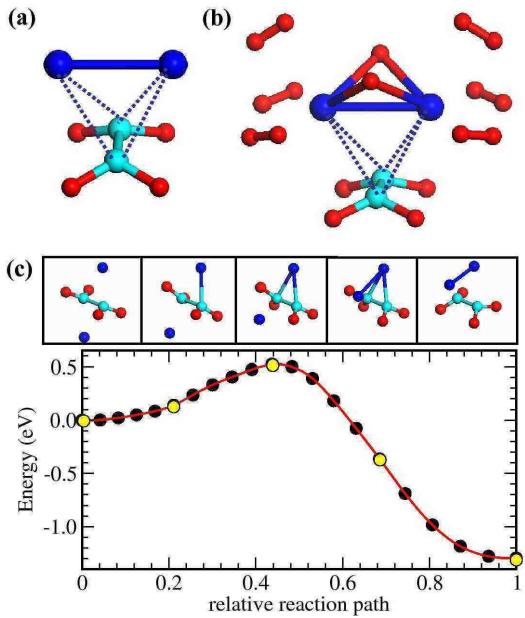}
\caption{
(Color online) (a) Bare C$_{2}$H$_{4}$Ti$_{2}$ dimer-perp complex. (b) The 
complex with seven H$_{2}$ absorbed. Note that there also exists other 
stable configurations that not discussed here. (c) The minimum energy path for the 
transition from C$_{2}$H$_{4}$Ti$_{2}$ sandwich configuration to dimer-perp configuration. The 
activation energy is about 0.55 eV. Note that regardless of which
isomer of C$_{2}$H$_{4}$Ti$_{2}$ we have, the resulting complex is able to 
bind multiple hydrogen molecules.
}
\label{rp_dimer}
\end{figure}

\begin{figure}
\includegraphics[width=7cm]{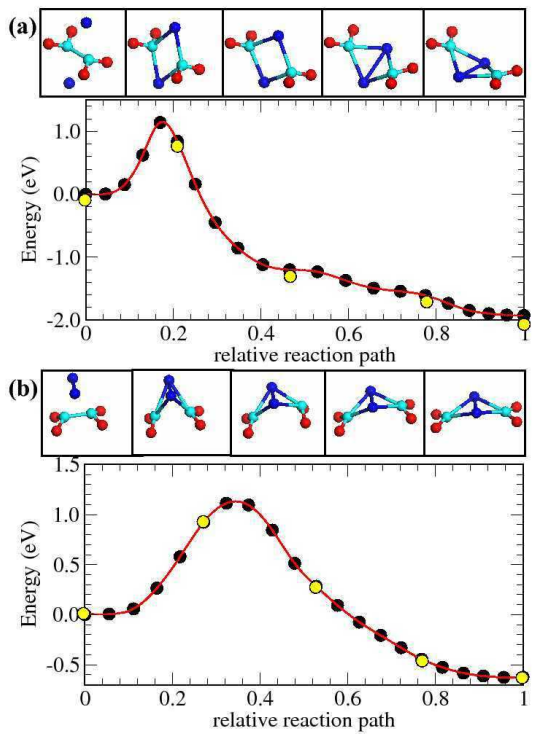}
\caption{
(Color online) The minimum energy paths for the transitions (a) from the C$_{2}$H$_{4}$Ti$_{2}$ 
sandwich to the dissociated configuration and (b) from
C$_{2}$H$_{4}$Ti$_{2}$ dimer-perp to the dissociated C$_{2}$H$_{4}$ 
configuration, respectively. In both cases, there are large energy barriers 
on the order of 1.1 eV.  
}
\label{rp_disso}
\end{figure}

The C$_{2}$H$_{4}$M$_{n}$ and C$_{2}$H$_{4}$M$_{n}$-H$_{x}$ complexes can 
have several isomeric structures. It is important to know the relative 
stabilities of these isomers and their implications for the hydrogen 
absorption properties. We thus studied the activation energies and reaction 
paths between different isomers of C$_{2}$H$_{4}$M$_{n}$ complexes. Here we 
discuss representative results on M=Ti.

We start with the C$_{2}$H$_{4}$Ti+H$_{2}$ complex and consider two possible 
structural transitions, which lead to lower energy configurations through 
the dissociation of an H$_{2}$ molecule over a Ti atom. In the first case, 
the H$_{2}$ molecule dissociates on top of the Ti atom. 
C$_{2}$H$_{4}$(Ti+H$_{2})$ and C$_{2}$H$_{4}$(TiH$_{2})$ are the reactant 
and product, respectively. Their relaxed structures correspond to the first 
and last images shown in the top panel of Fig. 6(a). The calculated minimum 
energy path for this process gives $\sim $0.25 eV barrier, which is small 
but still significant since the C$_{2}$H$_{4}$(Ti+H$_{2})$ configuration 
corresponds to a local energy minimum and possesses a H$_{2}$ binding energy 
of $\sim $0.3 eV. In the second case, the H$_{2}$ molecule is first 
dissociated over Ti and then one of the H atoms goes to carbon, forming a 
CH$_{3}$ group. The activation energy plot for this process is shown in Fig. 
6(b), indicating a very low barrier of only $\sim $0.15 eV. Once the product 
(i.e., the last image of the top panel of Fig. 6(b)) forms, the CCTi-bond 
angle is very soft, resulting in the zero-temperature structure shown in the 
inset, which has only 30 meV lower energy than the product. The final 
structure of the molecule (inset) is isostructural to the ``ethanol'' 
molecule and therefore we call it ``titanol''.

Since the titanol molecule is fairly easy to form, it is important to check 
if this new complex still possesses the high-capacity H$_{2}$ absorption 
property. In Fig. 7, we show several stable hydrogen absorption 
configurations on a titanol molecule. With only one H$_{2}$, it can be 
absorbed molecularly (Fig. 7(a)) with a bind energy of 0.3 eV or absorbed 
dissociatively (Fig. 7(b)), yielding a TiH$_{3}$ structure, which has a 
binding energy of about 1.0 eV/H$_{2}$. We expect that the dissociation 
process may have a similar barrier to that found in Fig. 6 (a). Importantly, 
the titanol molecule can bind up to five H$_{2}$ as molecules (Fig. 7(c)) 
with an average binding energy of $\sim $0.4 eV/H$_{2}$.

Next, we study the C$_{2}$H$_{4}$M$_{2}$ dimer structures. For Ti, the 
dimer-perp structure (Fig. 8(a)) has lower total energy than the isomeric 
sandwich structure (Fig. 1(b)) and dimer-par structure (Fig. 1(c)). Fig.8(c) 
shows the activation barrier for the transition from the sandwich 
configuration to the dimer-perp configuration. The activation energy is 
about 0.55 eV. Shown in Fig. 8 (b) is one of the stable configurations that 
we identified for the hydrogen absorption on the C$_{2}$H$_{4}$Ti$_{2}$ 
dimer-perp structure. Apparently, regardless which isomer of 
C$_{2}$H$_{4}$Ti$_{2}$ that we have, the complex is always able to bind 
multiple hydrogen molecules.

Finally, one may ask whether it is possible for the metal to catalyze and 
dissociate the C$_{2}$H$_{4}$ molecule (i.e., break the C=C bond), forming a 
more stable structure as shown in Fig. 1(f). Our calculations show that the 
activation energies for a sandwich to dissociated C$_{2}$H$_{4}$ (Fig. 9(a)) 
and a dimer-perp to dissociated C$_{2}$H$_{4}$ configurations (Fig. 9(b)) 
are both large, $\sim $1.1 eV. Thus it is very unlikely that the 
dissociation would happen under near ambient condition. Interestingly, we 
found that even the dissociated structure can still absorb multiple H$_{2}$, 
in which case, the system is somewhat similar to a Ti metallocarbohedryne 
(met-car) cluster\cite{Zhao:2006}$^{,}$\cite{Akman:2006}.

\section{Finite Temperature First-Principles MD Simulations}

In order to further test the stability of the C$_{2}$H$_{4}$M$_{n}$-H$_{x}$ 
complexes and the relative strength of different interactions (such as 
M-C$_{2}$H$_{4}$ versus M-H$_{2}$ interactions) and to identify possible 
reaction paths, we have carried out extensive first-principles MD 
simulations in the microcanonical ensemble 
(NVE)\cite{Marx:2000}$^{,}$\cite{Frenkel:1996}. The system is first 
optimized and then random initial velocities are generated to yield twice 
the target-temperature. When the system is in equilibrium, half of this 
energy goes to the potential and therefore the final temperature oscillates 
around the target temperature. We note that due to the small atomic mass of 
some elements (e.g., Li and H) in our system, it is essential to use a small 
MD time step such as 0.5 fs. Furthermore, convergence criteria for energy at 
each MD iteration should be very accurate (we used 10$^{-7}$ eV) in order to 
avoid total energy/temperature drift (i.e., change in the total 
energy/temperature as a function of simulation time). Since we are studying 
an isolated molecular complex in free space, it is also important that we 
eliminate the six degrees of freedom (i.e. three rotations and three 
translations) of the molecule. When this is not done, we observed that the 
input temperature goes to totally uniform translation/rotation of the 
molecules rather than populating the vibrational modes after 1-2 ps 
simulations. In our simulations, we fixed one of carbon atoms and then two 
components of position of the other carbon atom and one component of M atom 
position (which prevents the rotation of the molecule in the CCM-plane). In 
this way, the total degrees of freedom allowed in our simulation are $N_F= 3\times (N-6)$, as 
expected for an isolated molecule. The temperature of the system is defined 
as $ T(t) = \sum_{i}  m_i v_i^2/(2K_B N_F )$ where $i$ runs over the atoms of the complex and $k_{B}$ is Boltzman's 
constant. The relative fluctuation is of the order of $1/\sqrt{N_F}$. We also note that 
since our system is very small (i.e. about a dozen of atoms), it is 
basically a collection of a small number of harmonic oscillators and 
therefore temperature fluctuations are large. In fact trying to control 
system temperature through velocity 
scaling\cite{Marx:2000}$^{,}$\cite{Frenkel:1996} at small time 
interval does not work and yields wrong results. The microcanonical ensemble 
is thus the best for our purpose and as we shall see below it works well 
provided that a small time step is used and the total energy/force 
calculations are accurate enough. Here we present representative results on 
M=Li, Ti as examples.

\begin{figure}
\includegraphics[width=7cm]{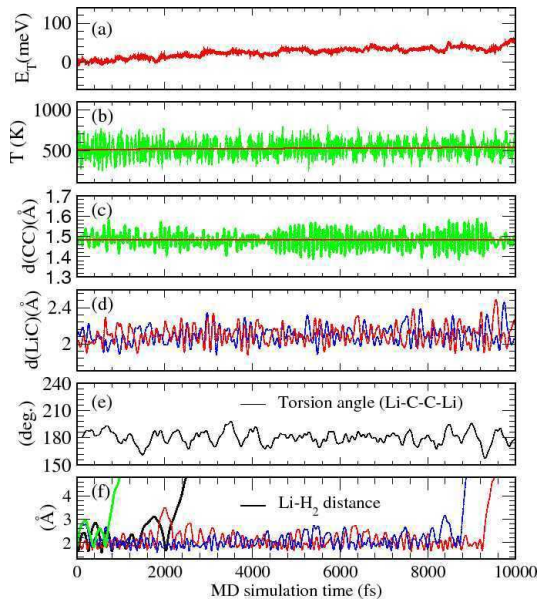}
\caption{
(Color online) First-principles MD results at 500K for the C$_{2}$H$_{4}$Li$_{2}$+4H$_{2}$ 
complex. Shown are the time evolution of various quantities, including total 
energy (a) and temperature (b) of the system, C-C (c) and Li-C (d) bond 
distances, and Li-C-C-Li torsion angle (e). The bottom panel (f) shows the 
distance between Li atom and the hydrogen center of mass, indicating 
successive desorption of H$_{2}$ molecules along the simulation.
}
\label{li_md}
\end{figure}

Our MD results for C$_{2}$H$_{4}$Li$_{2}$+4H$_{2}$ at 500 K are summarized 
in Fig. 10. The constant of motion plot shows only 50 meV drift in total 
energy over 10 ps simulation time, which causes a small temperature drift. 
The C-C and Li-C distances, shown in Fig. 10(c) and (d) respectively, 
indicate that the bare C$_{2}$H$_{4}$Li$_{2}$ molecule is stable at this 
temperature. The torsion angle Li-C-C-Li shows no sign of Li-dimer formation 
and oscillates around 180\r{ }. The bottom panel in Fig. 10 shows the 
distance between Li atoms and the center of mass of H$_{2}$ molecules, 
indicating the successive release of hydrogen molecules from the system. The 
first H$_{2}$ leaves the system around 400 fs. The fluctuations in the 
distances become very large at 2000 fs, resulting from the release of 
another hydrogen molecule. Around 8-10 ps, the other two hydrogen molecules 
also leave the system. Even though with 10 ps MD simulations, it is not 
possible to get reliable temperatures; the results are still very promising 
and suggest that the C$_{2}$H$_{4}$Li$_{2}$ system can stay intact at 500 K 
while it releases four hydrogen molecules.

We next studied the stability of C$_{2}$H$_{4}$Ti$_{n}$ system. We performed 
MD simulations up to 10 ps on C$_{2}$H$_{4}$Ti$_{2}$ (sandwich), 
C$_{2}$H$_{4}$Ti+H$_{2}$, C$_{2}$H$_{4}$(Ti+5H$_{2})_{2}$ (sandwich), and 
the titanol molecule (CH$_{3}$CH$_{2}$TiH) at 300 K and 500 K. In the 
simulations on the two metal sandwich systems, we did not observe any 
Ti-dimer formation. In the case for C$_{2}$H$_{4}$Ti+H$_{2}$, we did observe 
the spill-over effect, where the H$_{2}$ is dissociated over Ti and C and 
then Ti moved away with one hydrogen atom attached to it. This is 
essentially the titanol formation process that we discussed in the previous 
section.

The MD results for the C$_{2}$H$_{4}$(Ti+5H$_{2})_{2}$ system at 500 K are 
summarized in Fig. 11. During the 10 ps simulation time, both C-C and Ti-C 
bond distances oscillate around their equilibrium lengths without any 
indication of instability. Similarly, the Ti-C-C-Ti torsion angle also 
slowly oscillates around its equilibrium value of 180\r{ } and does not show 
any evidence for Ti-Ti dimer formation for which the torsion angle is 
supposed to be about 57\r{ }. Fig. 11(d) shows the number of H$_{2}$ 
molecules that are close to a Ti-atom (within a 2.2 {\AA} distance), showing 
that initially two H$_{2}$ molecules are released and then another H$_{2}$ 
molecule is released at around 2.4 ps. Above 6 ps, the number of H$_{2}$ 
fluctuates indicating that the distances are going beyond 2.2 {\AA} more 
often. Probably if we had run the MD simulation further, we would lose the 
remaining H$_{2}$ molecules that are attached to the Ti atoms.

\begin{figure}
\includegraphics[width=7cm]{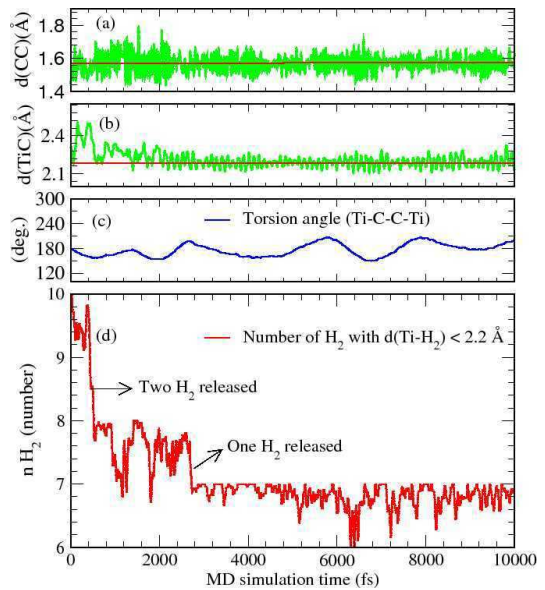}
\caption{
(Color online) First-principles MD results at 500K for the C$_{2}$H$_{4}$(Ti+5H$_{2})_{2}$ 
sandwich complex. Various quantities are shown, including C-C (a) and Ti-C 
(b) bond distances, and Ti-C-C-Ti torsion angle (c). The bottom panel (d) 
shows the number of H$_{2}$ molecules that are within 2.2 {\AA} 
of Ti atoms. It indicates successive desorption of H$_{2}$ molecules 
in the course of the simulation.
}
\label{ti_md}
\end{figure}

\begin{figure}
\includegraphics[width=7cm]{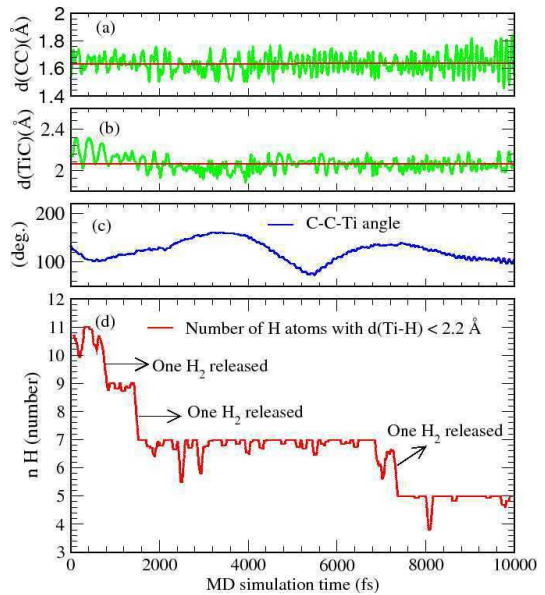} 
\caption{
(Color online) Various quantities (same as in Fig.11) obtained from MD 
simulations of Titanol+5H$_{2}$ system at 500 K.
}
\label{titanol_md}
\end{figure}

As a final example, in Fig. 12, we present results from a 10 ps MD run on 
titanol+5H$_{2}$ molecules at 500 K. The C-C and Ti-C distances indicate 
that the bare titanol molecule is stable at this temperature. The C-C-Ti 
angle shown in Fig. 12(c) indicates that C-C-Ti bond angle is very soft, 
exhibiting large amplitude motion. Around 5 ps, Ti actually goes to the 
middle of two carbon atoms, returning to our original C$_{2}$H$_{4}$Ti like 
configuration. As we discussed in the previous section, these two 
configurations are almost degenerate. The last panel shows the number of H 
atoms that are within 2.2 {\AA} of the Ti atom. Three successive 
constructive desorptions of H$_{2}$ molecule are evident.

In summary, our MD results discussed above on different systems indicate 
that the sandwich configuration of C$_{2}$H$_{4}$Ti$_{2}$ is quite stable 
and can bind H$_{2}$ molecules and then release them at elevated 
temperature. Similarly, C$_{2}$H$_{4}$Li$_{2}$ MD results also suggest that 
Li is another promising option even though the strength of the interactions 
is at the low side. Finally, thanks to MD simulations, we discovered a new 
configuration, titanol, which is derived from the C$_{2}$H$_{4}$Ti+H$_{2}$ 
system and capable of binding five H$_{2}$ molecules and then releasing them 
at high temperature without breaking down its structure.

\section{Conclusions}

Our conclusions are summarized as follows:

1. We showed that the C=C bond in ethylene can mimic the double bond in 
other carbon structures like C$_{60}$, in terms of binding metal atoms and 
the hydrogen absorption properties. The small system size of the M-ethylene 
complex allowed us to do very detailed studies such as long MD simulations 
and reaction path calculations, which were very difficult to perform 
otherwise. Most of the results that we found, such as H$_{2}$-dissociation 
and titanol formation, should be valid for other Ti-decorated 
nanostructures.

2. For light transition metals, we showed that the initial H$_{2}$ 
absorption could be either molecular with binding energy of $\sim $0.3 eV or 
it could be chemical by TiH$_{2}$ formation with a binding energy of $\sim 
$1.0 to 1.5 eV. However, there is a barrier of $\sim $0.25 eV for this 
process. Since the molecular H$_{2}$ has a binding energy of $\sim $0.3 eV, 
the dissociation could not be observed. Indeed, in our MD simulations, we 
did not see conversion from Ti+H$_{2}$ to TiH$_{2}$. Instead, we discovered 
that there is a very low energy barrier for the simultaneous dissociation of 
H$_{2}$ and formation of CH bonding (similar to spillover effect) through 
the Ti atom. For the case of C$_{2}$H$_{4}$Ti+H$_{2}$, this reaction yielded 
a new molecule which is isostructural to ethanol and can bind five hydrogen 
molecules with an average binding energy of $\sim $0.4 eV.

3. We showed that the sandwich configuration of C$_{2}$H$_{4}$M$_{2}$ is 
quite stable for both transition metals and Li. There are high energy 
barriers for the transition to dimer-configurations. Our 10 ps MD 
simulations did not show any evidence for dimerization.

4. From our results, it is clear that C$_{2}$H$_{4}$M$_{n}$ system could 
have a very rich phase diagram with different configurations. However, for 
all the isomer configurations that we have investigated, the complex is 
always able to bind multiple hydrogen molecules with high absorption 
capacity. Hence, these results suggest that co-deposition of 
transition/lithium metals with small organic molecules into nanopores of 
low-density materials could be a very promising direction for discovering 
new materials with better storage properties.

5. We note that there are many existing experimental studies of small 
organic molecules with transition metals in gas phase by mass spectroscopy. 
In these experiments, the metal atoms are obtained by laser evaporation of 
bulk metal and then condensed with mixture of Ar-and ethylene (or benzene) 
gas onto a cold substrate. In this way, it was possible to trap 
M$_{x}$(C$_{2}$H$_{4})_{y}$ types of complexes in an argon matrix and do 
spectroscopic experiments on them. We hope that our study will reenergize 
these studies with the focus on hydrogen absorption properties of these 
systems. It may be possible to use H$_{2}$ rather than Ar to prepare these 
clusters in an H$_{2}$-matrix. Such studies would be very important as a 
proof of concept and that should be the current emphasis.

\begin{acknowledgments}
We acknowledge partial DOE support
from  EERE grant DE-FC36-04GO14282 (WZ, TY) and and
BES Grant No. DE-FG02-98ER45701 (SC).
SC and ED acknowledge partial support 
from  T\"{U}B\.{I}TAK under Grant No. TBAG-104T536.
We thank J. Curtis and R. Cappelletti for fruitful discussions.
\end{acknowledgments}


\begin{thebibliography}{22}
\bibitem{See:2004} See the special issue Towards a Hydrogen Economy, by R. Coontz and B Hanson, Science \textbf{305}, 957 (2004).
\bibitem{Crabtree:2004} G. W. Crabtree, M.S. Dresselhaus and M.V. Buchanan, Physics Today, 39 (December 2004).
\bibitem{Zuttel:2003} A. Zuttel, Materials Today \textbf{6}, 24 (2003).
\bibitem{Yildirim:2005} T. Yildirim and S. Ciraci, Phys. Rev. Lett. \textbf{94}, 175501 (2005).
\bibitem{Yildirim:2006} T. Yildirim, J. Iniguez and S. Ciraci, Phys. Rev. B\textbf{72} 153403 (2005).
\bibitem{Zhao:2005} Y. Zhao, Y.-H. Kim, A. C. Dillon, M. J. Heben, and S. B. Zhang, Phys. Rev. Lett. \textbf{94}, 155504 (2005).
\bibitem{Dag:2005} S. Dag, Y. Ozturk, S. Ciraci and T. Yildirim, Phys. Rev. B \textbf{72}, 155404 (2005).
\bibitem{Kiran:2006} B. Kiran, A.K. Kandalam, and P. Jena, J. Chem. Phys. \textbf{124}, 224703 (2006).
\bibitem{Durgun:2006} E. Durgun, S. Ciraci, W. Zhou, and T. Yildirim, Phys. Rev. Lett. \textbf{97}, 226102 (2006).
\bibitem{Metal:2001} Metal Dihydrogen and Bond Complexes - Structure, Theory and Reactivity, edited by G.J. Kubas (Kluwer Academic/Plenum Pub. New York, 2001).
\bibitem{Manceron:1986} L. Manceron and L. Andrews, J. Phys. Chem. \textbf{90}, 4514 (1986).
\bibitem{Chen:1998} J. Chen, T. H. Wong, Y. C. Cheng, K. Montgomery, and P. D. Kleiber, J. Chem. Phys. \textbf{108}, (1998).
\bibitem{Manceron:1989} L. Manceron and L. Andrews, J. Phys. Chem. \textbf{93}, 2964 (1989).
\bibitem{Ozin:1978} G. A. Ozin, W. J. Power, T. H. Upton, and W. A. Goddard III, J. Am. Chem. Soc. \textbf{100}, 4750 (1978).
\bibitem{Alikhani:1996} M. E. Alikhani and Y. Bouteiller, J. Phys. Chem. \textbf{100}, 16092 (1996).
\bibitem{Sodupe:1992} M. Sodupe, C. W. Bauschlicher, S. R langhoff, and H. Partridge, J. Phys. Chem. \textbf{96}, 2118 (1992).
\bibitem{Blomberg:1992} M. R. A. Blomberg, P. E. M. Siegbahn, and M. Svensson, J. Phys. Chem.\textbf{ 96}, 9794 (1992).
\bibitem{Papai:1993} I. Papai, J. Mink, R. Fournier and D. R. Salahub, J. Phys. Chem. \textbf{97}, 9986 (1993).
\bibitem{Baroni:1} S. Baroni, A. Dal Corso, S. de Gironcoli, and P. Giannozzi, http://www.pwscf.org.
\bibitem{Mills:1994} G. Mills and H. Jansson, Phys. Rev. Lett. \textbf{72}, 1124 (1994).
\bibitem{Henkelman:2000} G. Henkelman and H. Jansson, J. Chem. Phys. \textbf{133}, 9978 (2000).
\bibitem{Marx:2000} D. Marx and J. Hutter, \textit{Ab-initia Molecular Dyanimcs: Theory and Implementation}, in \textit{Modern Methods and Algorithms of Quantum Chemistry} (p. 301-449), Editor. J. Grotendorst (NIC, FZ Julich 2000).
\bibitem{Frenkel:1996} D. Frenkel and B. Smith, \textit{Understanding Molecular Simulation} (Acaemic Press: New York, 1996).
\bibitem{Durgun:2003} E. Durgun, S. Dag, V. K. Bagci, O. G\"{u}lseren, T. Yildirim and S. Ciraci, Phys. Rev. B \textbf{67}, 201401(R) (2003).
\bibitem{Durgun:2004} E. Durgun, S. Dag, S. Ciraci and O. G\"{u}lseren, J. Phys. Chem. B \textbf{108}, 575 (2004).
\bibitem{Georges:1999} R. Georges, M. Bach and M. Herman, Mole. Phys. \textbf{97}, 279 (1999).
\bibitem{Zhao:2006} Y. Zhao, A. C. Dillon, Y.-H. Kim, M. J. Heben, and S. B. Zhang, Chem. Phys. Lett. \textbf{425}, 273 (2006).
\bibitem{Akman:2006} N. Akman, E. Durgun, T. Yildirim and S. Ciraci, J. Phys: Condens. Matter, J. Phys.: Condens. Matter \textbf{18}, 9509 (2006).
\end{thebibliography}
\end{document}